\documentclass{aa}
\usepackage{graphicx}
\newcommand{\figwidth}{90mm}
\newcommand{\be}{\begin{equation}}
\newcommand{\ee}{\end{equation}}
\newcommand{\hsl}{h_{sl}}
\begin{document}
   \title{Dome C site testing: surface layer, free atmosphere seeing and isoplanatic angle statistics.}

   \subtitle{}

    \author{E. Aristidi \and E. Fossat \and A. Agabi \and D. M\'ekarnia \and F. Jeanneaux \and E. Bondoux \and Z. Challita \and A. Ziad \and J. Vernin \and H. Trinquet}

   \offprints{E. Aristidi, \email{aristidi@unice.fr}}

   \institute{Laboratoire Hippolyte Fizeau,
              Universit\'e de Nice Sophia Antipolis, Parc Valrose,
          06108 Nice Cedex 2, France
 }
   \date{Received: / Accepted:}

   \abstract{This paper analyses 3$\frac{1}{2}$ years of site testing data obtained at Dome C, Antarctica, based on measurements obtained with three DIMMs located at three different elevations. Basic statistics of the seeing and the isoplanatic angle are given, as well as the characteristic time of temporal fluctuations of these two parameters, which we found to around 30 minutes at 8 m. The 3 DIMMs are exploited as a profiler of the surface layer, and provide a robust estimation of its statistical properties. It appears to have a very sharp upper limit (less than 1 m). The fraction of time spent by each telescope above the top of the surface layer permits us to deduce a median height of between 23 m and 27~m. The comparison of the different data sets led us to infer the statistical properties of the free atmosphere seeing, with a median value of 0.36 arcsec. The $C_n^2$ profile inside the surface layer is also deduced from the seeing data obtained during the fraction of time spent by the 3 telescopes inside this turbulence. Statistically, the surface layer, except during the 3-month summer season, contributes to 95 percent of the total turbulence from the surface level, thus confirming the exceptional quality of the site above it.

   }
\titlerunning{Dome C site testing}
   \maketitle
%
%________________________________________________________________

\section{Introduction}

The Italo-French Concordia station located on the Dome C site of the Antarctic plateau has been attracting more and more interest from the astronomical community since its official opening in 2005. The dry, cold and clean air as well as the absence of aerosols and light pollution are obvious advantages for infrared astronomy. The long nights and days are well suited for long term monitoring programs. The vertical repartition of the turbulence is regarded as extremely interesting for high angular resolution imaging. 

It is now well known that the surface seeing on the Antarctic plateau is quite poor. It has been measured at about 1.7 arcsec in the visible at the South Pole (Marks et al. 1999; Travouillon et al. 2003). However, microthermal radio soundings by balloon-borne sensors estimated a seeing that would be around 0.3 arcsec above a turbulent surface layer  (SL), measured to be of 200m thickness at the South Pole. Because of the geographical situation, this SL is expected to be much thinner at Concordia (see for instance Swain at Gallee 2006). Indeed, the same radio soundings made at Concordia during the first winter-over season in 2005 revealed a median thickness of 32m and a seeing likely to be of the order of 0.4 arcsec or better above it, while it is not significantly better than at the South Pole at surface level (Trinquet et al., 2008a).

The important parameters to be explored in more detail are the statistical behaviour of the SL thickness, the seeing above it (noted hereafter as above surface layer (ASL) to avoid confusion with the so-called free atmosphere that starts a few hundred meters higher), their possible seasonal dependence, and the isoplanatic angle, again with its seasonal dependence. The summer seeing parameters have  been extensively studied (Aristidi et al. 2005). A first estimation of the free atmosphere seeing was provided by a SODAR+MASS analysis by Lawrence et al. (2004). The exceptionally small values of the turbulence outer scale have been measured by the GSM (Ziad et al. 2008), and also deduced from winter radio soundings (Trinquet et al. 2008b), while the isopistonic angle has been estimated by Elhalkouj et al. (2008) who demonstrated that a 3~m telescope at Dome C would have the same isopistonic angle as the Keck in Hawaii.

This paper shows that 3 DIMMs operated at 3 different altitudes can be exploited like a turbulence profiler of the SL, and can give access to both its geometrical statistics and its typical $C_n^2$ content and profile. Given the fraction of time spent by these DIMMs above the upper limit of the SL, the free atmosphere seeing is also accessible with a robust statistical significance. The time dependence of the seeing parameters, on short scales as well as the possible seasonal variations, is also adressed. A surprising result is the very large fraction (nearly 95\%) of the total turbulence found to be inside this thin SL, making the site conditions very useful just above it.

%
%________________________________________________________________

\section{Results}
\subsection{Seeing statistics}
The seeing measurements exploited here consist of 3 different time series. The main one was obtained by the DIMM located on the wooden Concordiastro platform (about 8 m height) during the first summer campaign (2003-2004) and then nearly continuously since December 2004, thus covering more than 3 and a half years of data in 320000 measurements (a seeing point is computed from a set of 2~minutes of data). The overal duty cycle since December 2004 is of 36\%. A fraction of the missing measurements of around 10\% (Mosser \& Aristidi 2007) is due to adverse weather, the rest to technical difficulties. However, the longest permanent gap during these nearly continuous 3 and a half years is less than 18 days. Each of the 14 consecutive seasons covered by this data set has enough measurements to undertake significant statistical studies. A second time series comes from the ground-based DIMMs operated as a GSM instrument (Ziad et al. 2008). It contains 227000 measurements covering the period from December 2004 to April 2008. The duty cycle during this period was only of 25\%, including a gap longer than 6 months during the winter 2006 (fatal technical failure). During the period July - October 2005, one of the telescopes of the GSM experiment was moved to the roof of the so-called calm building of Concordia and provided 24000 measurements at an elevation of 20 m above the ground. GSM is composed of 2 identical DIMMs operated at about 3 m above the plateau. For 3 months in the winter and spring periods, it was then possible to obtain simultaneous values of the seeing from 3 different heights: 3 m, 8 m and 20 m. The 20 m monitor was however located at a horizontal distance of 300 m from the two others, and is suspected to be contamined by local turbulence generated by the building, even though it was exposed in the prevaling wind direction. 

\begin{figure}
\includegraphics[width=\figwidth]{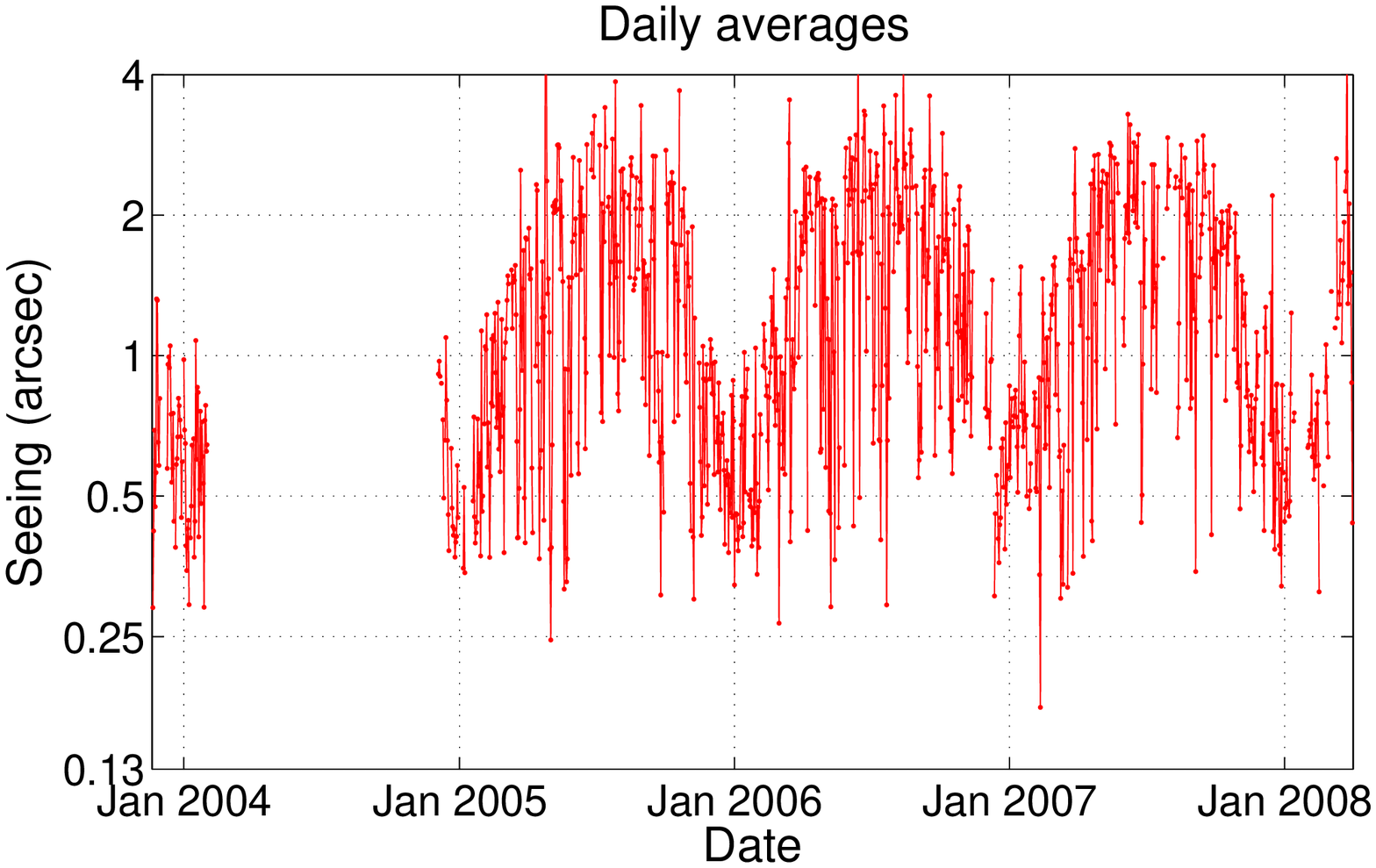} 
\includegraphics[width=\figwidth]{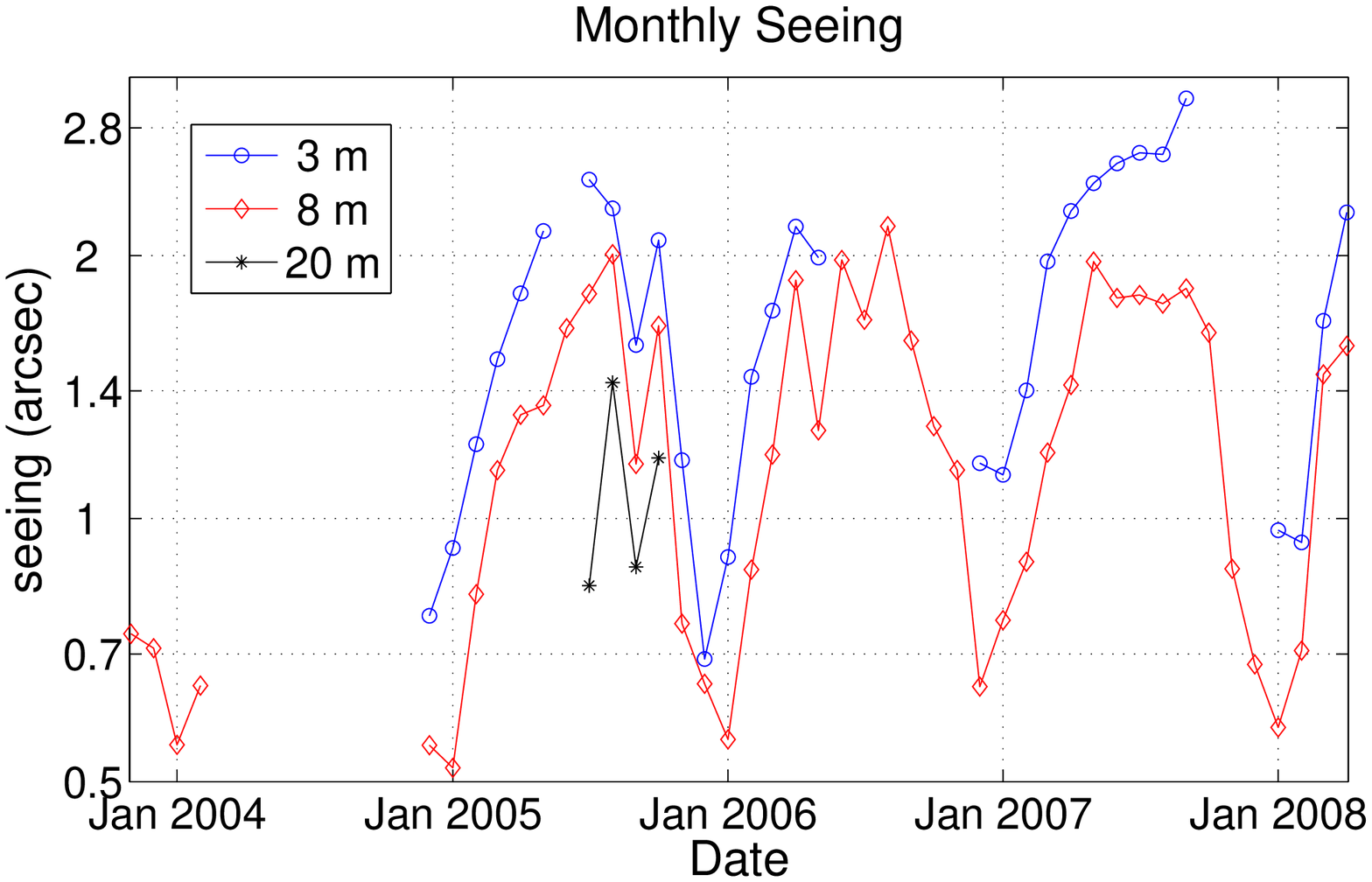}
\caption{Top: daily average seeing at an elevation of 8 m as a function of time. Bottom: monthly seeing for the three elevations 3 m, 8 m and 20 m respectively. Seeing scale is logarithmic.}
\label{fig:dailyseeing8m}
\end{figure}

Daily and monthly averages of the seeing are plotted as a function of time in Fig.~\ref{fig:dailyseeing8m}. We see a strong dependence of the seeing with the season. The best conditions are observed in the summer with a median of 0.54\arcsec at an elevation of 8~m, then the seeing degrades to values around 2$\arcsec$ in August. This behaviour was noticed since the first winterover (Agabi et al. 2006; Trinquet et al. 2008a), and seems to occur every year. It was explained by the occurrence, in winter, of a strongly turbulent SL whose thickness is about 30~m (Trinquet et al. 2008a). Seeing values at the three available elevations are summarized in Table~\ref{tabl:seeingstat}.

% Figure 1

\begin{table*}
\begin{center}
\begin{tabular}{lrrrrrrr}\hline
Elevation							 		& & 3 m & & & 8 m & & 20 m \\ 
                          & summer & winter & total & summer & winter & total & \\ \hline
Mean        & 1.06 & 2.51 & 1.83    & 0.69 & 1.72 & 1.23  & 1.10 \\
Median      & 0.95 & 2.37 & 1.67    & 0.57 & 1.65 & 0.98  & 0.84  \\
P$_{75}$    & 1.32 & 2.98 & 2.38    & 0.86 & 2.32 & 1.69  & 1.55 \\
P$_{25}$    & 0.70 & 1.86 & 1.06    & 0.40 & 0.83 & 0.52  & 0.43 \\
Max         & 4.76 & 9.26 & 9.78    & 7.63 & 9.09 & 15.05 & 8.20 \\
Min         & 0.03 & 0.24 & 0.03    & 0.03 & 0.13 & 0.03  & 0.13  \\ \hline
\end{tabular}
\end{center}
\caption{Global seeing statistics for the three available DIMM data. P$_{75}$ and P$_{25}$ are the 75\% and 25\% percentiles. The data at an elevation 20 m are limited to the period from July to October 2005.}\label{tabl:seeingstat}
\end{table*}

\begin{figure}
\includegraphics[width=\figwidth]{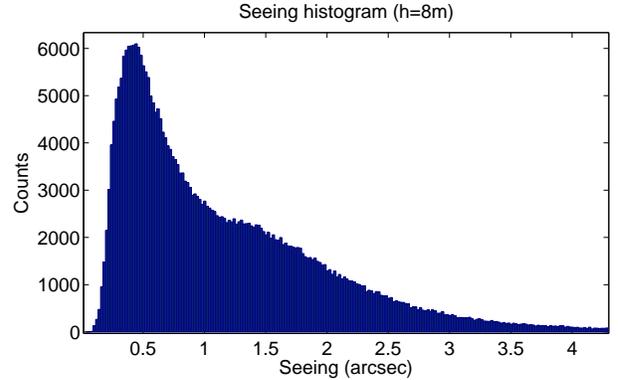}
\caption{The histogram of 3 and a half years of DIMM seeing data recorded at an elevation of 8m on the Concordiastro platform reveals two different regimes corresponding to good and bad seeing.}
\label{fig:histoseeing8m}
\end{figure}

\begin{figure}
\includegraphics[width=\figwidth]{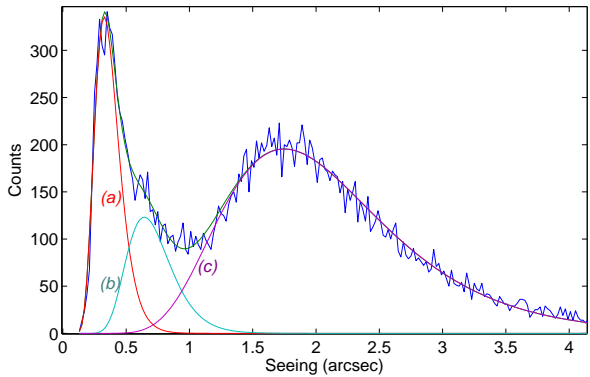}
\caption{The fit of 3 log-normal curves of the winter 2006 histogram of the 8m DIMM. It clearly distinguishes 3 different situations : (a): an excellent seeing distribution, corresponding to situations where the telescope is above the SL. (c): a poor seeing distribution (the telescope is totally embedded inside the SL). (b): the distribution of less frequent intermediate events. 
}
\label{fig:histoseeingwinterfit}
\end{figure}

The large volume of the available data set encourages us to use the basic tools of statistics, the simplest one being histograms. A first histogram looked at is the longest time series, including all seasons and obtained at 8m height (Figure~\ref{fig:histoseeing8m}). It shows a bi-modal distribution, each one of the two peaks having a log-normal shape. However, this all-season histogram mix the very different conditions that prevail in summer and winter. It is then highly preferable to analyse the histograms season by season, this being permitted by the large size of the data sets. The seasons are called Summer (when the Sun never set, Nov. 1st - Feb. 4th), Winter (when it never rises, May 4th - August 11th), the other two being Autumn  (Feb. 4th - May 4th) and Spring (August 11th - Nov. 1st). Figure \ref{fig:histoseeingwinterfit} is an example of a winter histogram (2006). The bi-modal appearance is more striking, with a sharp peak of excellent seeing centered at 0.3 arcsec, and a broader peak of poor seeing located around 1.7 arcsec. As this is a one-season histogram, the bi-modal appearance cannot be interpreted as possible evidence for two different regimes of summer and winter. It seems to indicate that the telescope stands either inside or outside the SL. This indicates that the upper limit of the SL must be rather sharp. 
We attempted to model the histograms by a sum of two log-normal functions, but it failed in the dip between the two curves. A sum of three log-normal functions was required to obtain the good fit that is visible here. Figure 4, for comparison, is an autumn histogram, from 2007. It shows a very similar general behaviour: a poor seeing distribution, now peaked around 1.3  arcsec, a good seeing distribution still peaked at 0.3 arcsec, and a similar third component. The fitted functions are defined by  
\be
\begin{array}{ll}g(x)= & \displaystyle a_1 \exp\left(\frac{x-a_2}{a_3}\right)^2 + b_1 \exp\left(\frac{x-b_2}{b_3}\right)^2 +\\ \ \\
 & \displaystyle   c_1 \exp\left(\frac{x-c_2}{c_3}\right)^2 
\end{array}
\label{eq:fit}
\ee
where $x$ is the natural logarithm of the seeing. The addition of these histograms assumes that the SL and the ASL atmosphere are statistically independent. The interpretation of these curves is that most of the time, the telescope is embedded inside the SL (curve c), while for a smaller fraction of the time, it is totally outside the SL (curve a). There remains an intermediate situation (curve b) that can correspond to two cases: (i) the SL is not unique, but contains a second (and weaker) layer above the telescope (such situations were shown by the radio soundings, see Fig. \ref{fig:ballprof} for examples), (ii) the SL upper limit is just in front of the entrance window of the telescope and moves slightly up and down during the two-minute integration time of one seeing measurement. 

% Figure 2

% Figure 3

% Figure 4
\begin{figure}
\includegraphics[width=\figwidth]{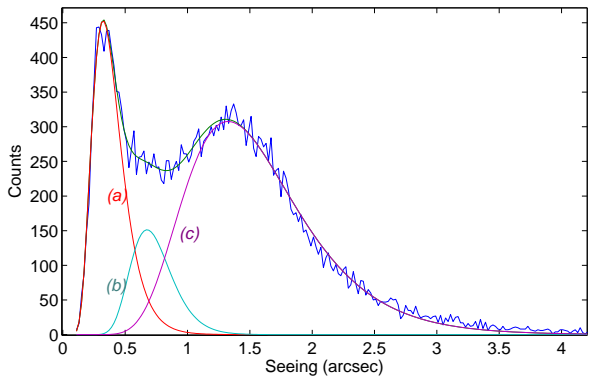}
\caption{An autumn histogram showing a situation very similar to the winter one displayed in Fig.~\ref{fig:histoseeingwinterfit}, the only difference being the weaker turbulent energy inside the SL (the maximum of the curve (c) is centered on a smaller value). 
}
\label{fig:histoseeing20m8m}
\end{figure}

%
%________________________________________________________________

\subsection{A statistical model of the surface layer}
As the three telescopes have been operated at 3 different altitudes the histograms and their modeling by the sum of log-normal functions give access to a 3-level vertical profile of the turbulence strength $C_n^2$ inside the SL, season by season. Indeed, curve (a) shows almost no displacement all year round and can then be regarded as representing the seeing above the SL : it provides a robust estimation of the ASL seeing probability density function (PDF); this will be discussed in Section \ref{par:faseeing}. Its surface gives the probability of the SL of being thinner than the telescope height (respectively 18\% and 24\% in the examples shown in Figures  \ref{fig:histoseeingwinterfit} and \ref{fig:histoseeing20m8m}). Similarly, the curve (c) gives the probability of the main part of the SL being thicker than the telescope height (respectively 70\% and 65\% in the same examples).

The histogram of the data set of the 20 m high DIMM gives a probability of about 45\% of it being outside the SL. It is  interesting to compare it with the histograms of the data obtained simultaneously at 8 m (Fig \ref{fig:histoseeing20m8m}). Though the telescopes were not absolutely synchronized and each data set exhibits different gaps, these measurements were made during the same 3 and a half months. 

The left peak of the two histograms of Fig. \ref{fig:histoseeing20m8m} is expected to represent the same function, i.e. the ASL seeing PDF, as will be discussed in section \ref{par:faseeing}. There should only be a difference in the weight of these functions regarding the total seeing distribution (18\% at 8 m, 45\% at 20 m). Figure \ref{fig:ratiohistoseeing} shows the ratio  (20m/8m) of the two normalized histograms. In its left part (i.e. for seeing values below 0.5 arcsec) it should tend towards the ratio of the two ASL PDFs, i.e. the ratio of two identical functions, which is a constant. This constant is the ratio of 45\% to 18\%, i.e. 2.5. However this expected saturation of the graph for small values (represented with the dashed line in Figure \ref{fig:ratiohistoseeing}) is not observed. Instead, we have a depletion for seeings between 0.1 and 0.3 arcsec, and too many points in the range 0.3 -- 0.45 arcsec. The extreme values, around 0.1 arcsec, are missing at 20 m while they sometimes do appear at 8 m. Very likely, the seeing measurements made by this DIMM situated on the roof of the building suffer from the large amplitude mechanical vibrations created by this situation. 
%They seem not to give a reliable measurement in these ultimate conditions of exceptional seeing. The behaviour of the ratio curve can be understood by assuming that the extreme values of 0.1 to 0.25 arcsec have been slightly damaged to become 0.3 to 0.45 arcsec, thus increasing the ratio in this range and decreasing it below 0.3. 
Extreme seeing values of 0.1\arcsec to 0.25\arcsec are biased, and measured in the range 0.3\arcsec -- 0.45\arcsec. However, the mean value of this ratio between 0.1 and 0.45 arcsec remains equal to 2.5. This supports the robustness of the DIMM measurements, that appears to be insensitive to the building vibrations if the seeing is not better than 0.3 or 0.4 arcsec.

The third data set,  excluding summers, from the ground-based DIMM is shown in Figure \ref{fig:histoseeing3m}. The mean seeing is poor, almost 2 arcsecs on average, but two bumps of extremely good (curve a) and very good (curve b) seeing, around 0.3 and around 0.65 arcsec are seen. They correspond respectively to probabilities of 2\% and 5\%. Curve a shows a small fraction of ASL seeing, when the SL is totally confined below 3m or vanishes, while the other one shows the intermediate situation. This is further confirmation of the sharpness of the upper edge of the SL, when its total thickness is not higher than 2 or 3 meters.

\begin{table*}
\begin{tabular}{c c c c c c c c}\\ \hline  \hline
season&Ntotal&Na/N&Nb/N&Nc/N&seeing a&seeing b&seeing c\\  \hline

8 m DIMM&&(\%)&(\%)&(\%)&(\arcsec)&(\arcsec)&(\arcsec)\\

winter 2005&29000&13.4&22.1&64.6&0.26&0.46&1.68\\
winter 2006&24000&17.75&11.70&70.55&0.32&0.64&1.75\\
winter 2007&17000&19.00&4.4&76.60&0.43&0.95&1.79\\

winter averages&70000&16.24&14.36&69.40&0.33&0.54&1.73\\

autumn 2005&25000&16.18&21.20&62.62&0.27&0.55&1.22\\
autumn 2006&24000&15.00&9.10&75.90&0.32&0.66&1.05\\
autumn 2007&29000&24.64&11.00&64.36&0.32&0.67&1.32\\

autumn averages&78000&19.00&13.70&67.30&0.30&0.61&1.20\\

spring 2005&16000&0.68&27&72.32&0.25&0.42&1.49\\
spring 2005$^\dag$&16000&27& &73&0.41&&1.49\\
spring 2006&14000&11.87&13.28&74.84&0.37&0.69&1.58\\
spring 2007&14000&6.52&45.29&48.17&0.27&1.08&1.70\\

spring averages&44000&6.10&28.54&65.80&0.29&&1.58\\

general averages&&17.3 $\pm$ 2.3&&67.2 $\pm$ 3.2&0.32&&1.48\\  \hline

ground DIMM (3m)&114590&&&&&&\\
23/2 - 19/5  2005&&&&&&&\\
21/7 - 29/9  2005& &2.0&5.2&92.8&0.32&0.74&1.914\\
19/3 - 27/8  2007&&&&&&&\\  \hline

roof DIMM  (20 m)&23913&&&&&&\\
23/7 - 31/10 2005&&&&&&&\\ 
3 components&&15.7&29.3&55&0.298&0.42&1.165\\
2 components&&45&&55&0.37&&1.16\\ \hline \hline
\end{tabular}\\
{\scriptsize}$\dag$: fit with 2 log-normal components
\caption{Table of fitted parameters relevant to Eq. \ref{eq:fit} for the 3 DIMMs at 3, 8 and 20m and for each season and/or each year. Columns labelled as ``$N_a/N$'', ``$N_b/N$'' and ``$N_c/N$'' are the integrals under the 3 fitted log-normal functions in \%. Columns ``seeing a'', ``seeing b'' and ``seeing c'' are the abcissa of the maximum of the 3 fitted functions.}
\label{table:parfit}
\end{table*}

Table \ref{table:parfit} gives the fits for all individual seasons and for the 3 telescopes. In most cases, they are the result of the 3-curve fit described above. In a few situations, a 2-curve fit was sufficient.

%
%________________________________________________________________

\subsection{The surface layer thickness}
Table \ref{table:parfit} shows that the SL significantly changes with the season, being stronger in winter, weaker in autumn, and intermediate in Spring. We already know that it is definitely much weaker in summer, even disappearing totally every day at 5 p.m. However it seems that only its turbulent energy changes, and not its geometry. The small variations of the mean probabilities (integral of curve c) between Autumn, Winter and Spring are not statistically significant.  All these probabilities can then be plotted as a function of altitude (Figure \ref{fig:prob_ins_bl}). Assuming that sometimes the SL contains a second component above the first main one, this plot shows two probabilities for each value $h_{sl}$ of the height: the highest means all the content of the SL is less than $h_{sl}$, while the lowest means only the main, most energetic part of this SL is less than $h_{sl}$. The 3-curve fits on the DIMM data sets provide these numbers at 3, 8 and 20m. At higher altitudes, a hint can be obtained from the 2005 winter radio soundings (Trinquet et al, 2008a). Figure \ref{fig:ballprof} shows the  32 available SL profiles with a vertical resolution of the order of 5 to 10 meters (depending on the vertical speed of the balloons). It shows a turbulent profile drastically changing from sample to sample, and ending  very sharply, at a mean height of about 30 meters, and sometimes including a second bump. We estimate the two probabilities from the profiles up to an elevation of 50~m. The error bars of the DIMM measurements come from both the scatter of individual seasonal histograms at 8m, and from the statistical error. The error bars for the balloon points assume that all the radio soundings are independent, with a Poisson probability (the mean equals the variance). From this figure, the mean and the median SL thickness are both of the order of 25 meters.

\begin{figure}
\parbox{90mm}{\hskip -2mm \includegraphics[width=47mm]{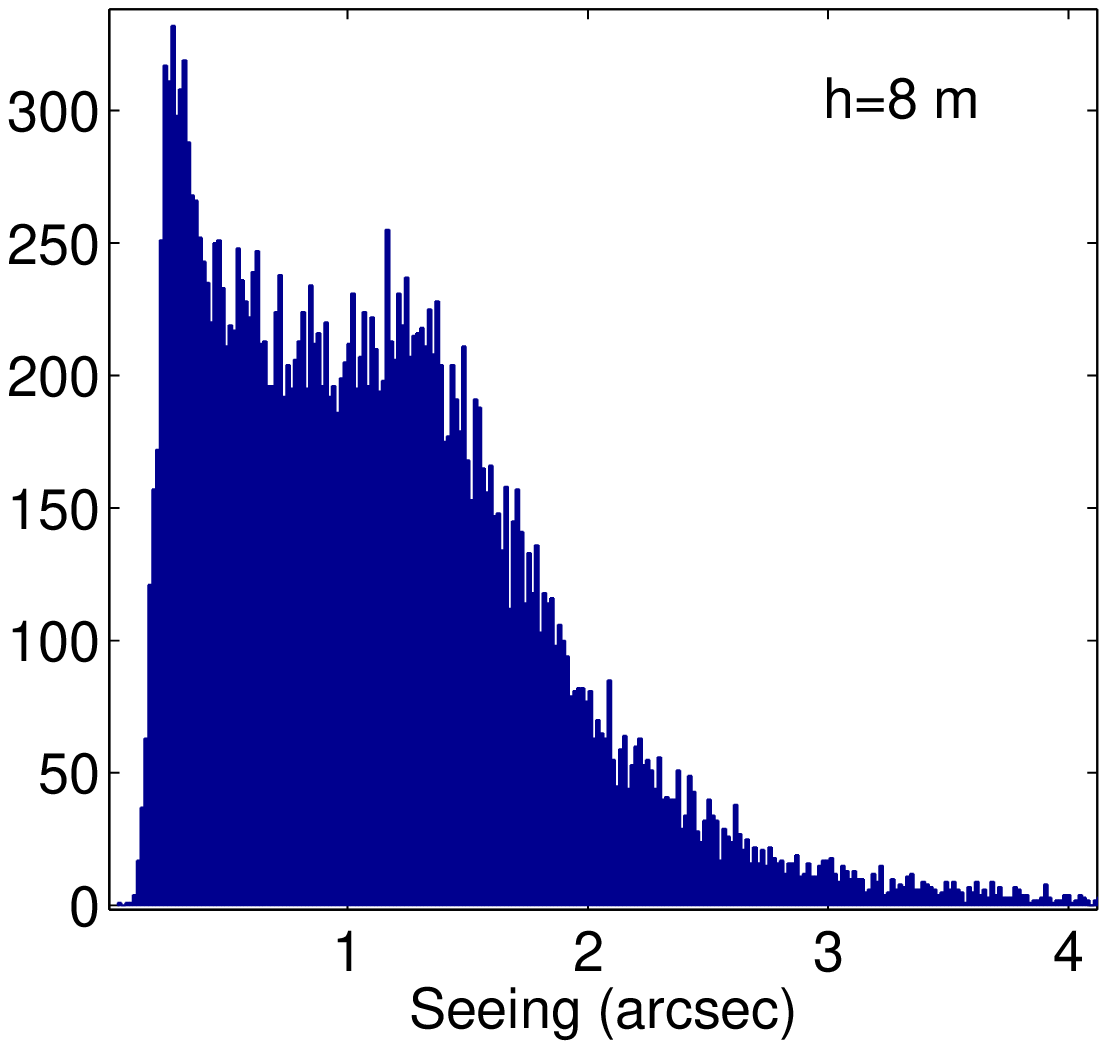}\nolinebreak
\hskip -1 mm  \includegraphics[width=47mm]{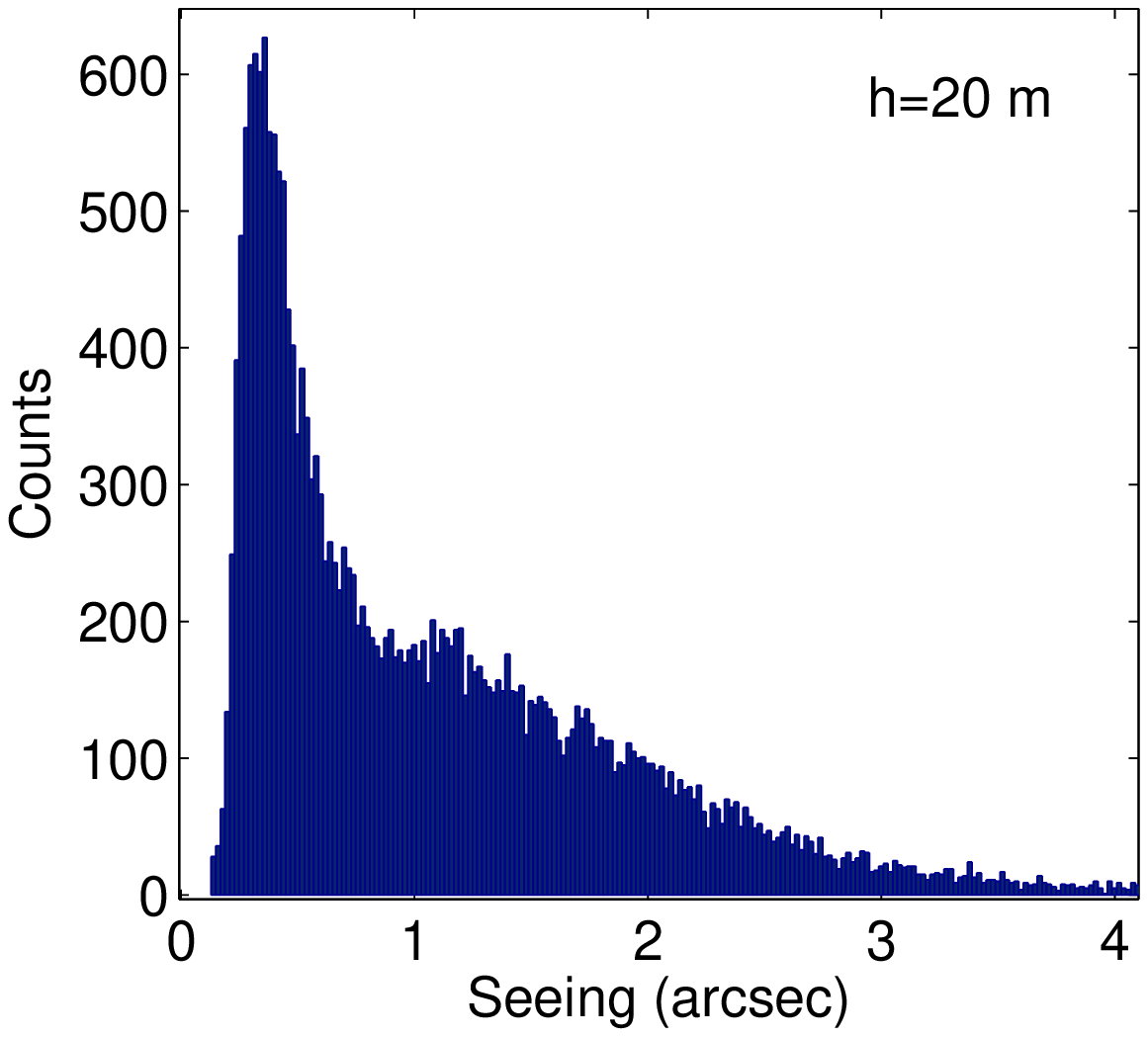}}
\caption{Histograms of the seeing obtained during the same 4 months in 2005 at 8 m and 20 m showing how much the seeing improves in this 12 m vertical interval. Quantitatively, they provide the probabilities of beeing above the SL at these two altitudes.}
\label{fig:histoseeing20m8m}
\end{figure}

\begin{figure}
\includegraphics[width=\figwidth]{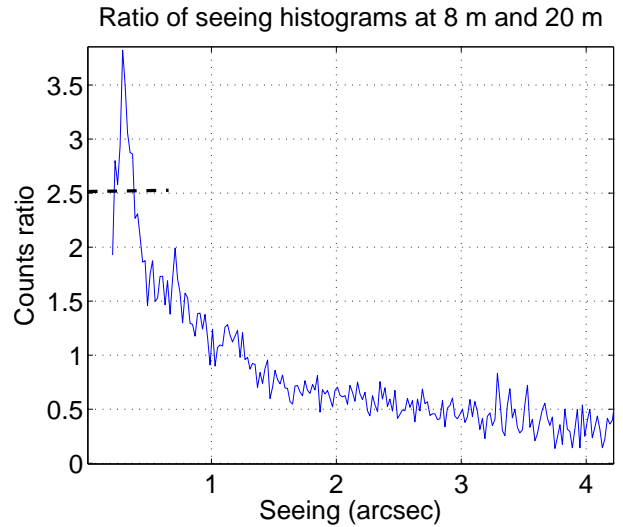}
\caption{Ratio of the two histograms of Fig. \ref{fig:histoseeing20m8m}. The dashed line at count ratio=2.5 corresponds to the expected horizontal asymptote of the curve when the seeing tends to zero (see text).}
\label{fig:ratiohistoseeing}
\end{figure}

\begin{figure}
\includegraphics[width=\figwidth]{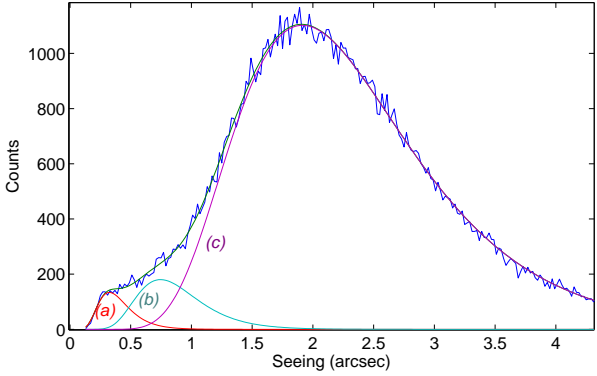}
\caption{Histogram of the seeing data taken by the 3-m DIMM, as usual excluding summers. As for Fig. \ref{fig:histoseeingwinterfit}, curves  (a), (b), (c) represent the three log-normal fits. The mean seeing is quite poor (close to 2 arcsec), but the excellent and intermediate seeing situations are still visible, even with a telescope mount placed on the snow surface,  with probabilities of about 2\% and 5\%.
}
\label{fig:histoseeing3m}
\end{figure}

\begin{figure}
\includegraphics[width=\figwidth]{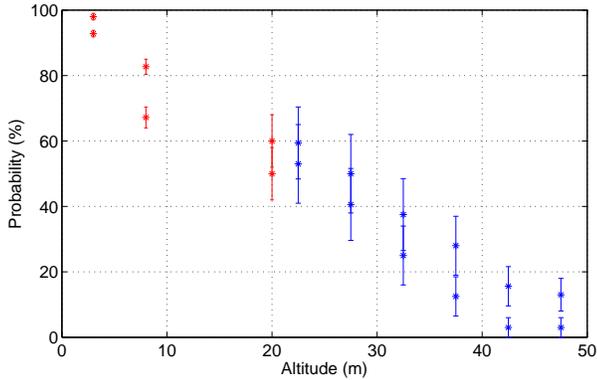}
\caption{Probabilities of being inside the SL as a function of altitude above the snow surface. See text (section 2.3) for details. 
}
\label{fig:prob_ins_bl}
\end{figure}

% figure 9
\begin{figure*}
{\begin{center} \includegraphics[width=150mm]{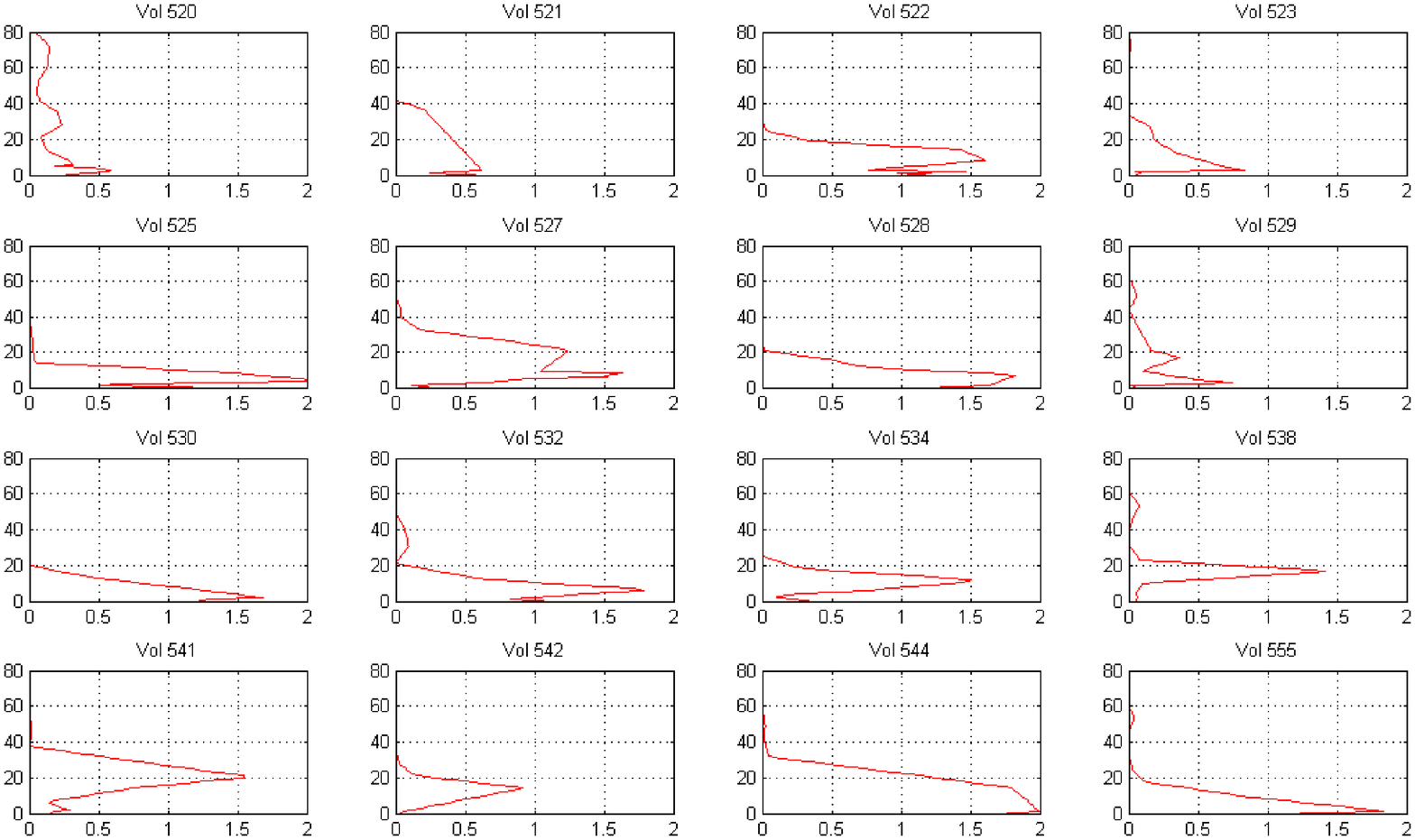}\\ \includegraphics[width=150mm]{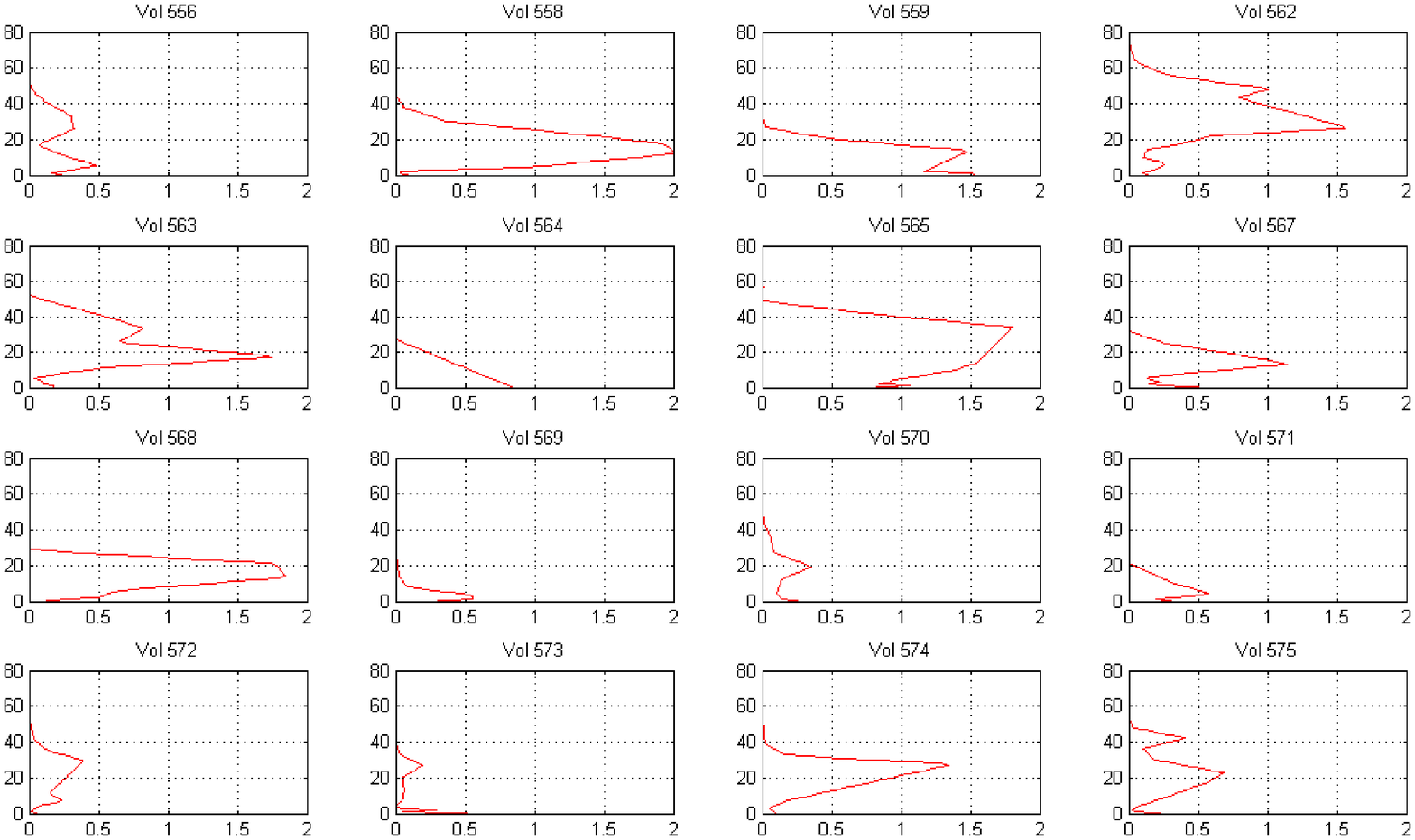}\end{center}}
\caption{ 32 samples of the SL $C_n^2$ vertical distribution in the first 80 m above the ground, measured during winter 2005 by balloon radio soundings. The horizontal axis is $C_n^2$ in units of $10^{-13}$ m$^{-2/3}$ and the vertical axis is the altitude over the ground in meters. Note that the $C_n^2$ is displayed in linear scale (it is often shown in log scale).}
\label{fig:ballprof}
\end{figure*}

%
%________________________________________________________________

\subsection{The $C_n^2$ vertical distribution inside the surface layer}
The $C_n^2$ vertical distribution inside the surface layer can be inferred from curves c. The subsets of data that provide these ``poor seeing'' parts of the histograms correspond to all situations when the SL exceeds the altitude of each telescope. The seeing produced both by the upper part of the SL, above the telescope located at the altitude $h_t$, and the rest of the free atmosphere, is defined by:
\be
\epsilon(h_t)= 5.25\: \lambda^{-1/5}\: S(h_t)^{3/5}
\label{eq:seeingsh}
\ee
where $S(h_t)$ is the integral upwards of the local turbulent strength $C_n^2(h)$  and $\epsilon(h_t)$ is in radians.
\be
S(h_t) =  \int_{h_t}^\infty C_n^2(h)\: dh
\label{eq:shdef}
\ee
The three seeing data sets corresponding to the period Jul--Oct 2005 were binned into 15~mn intervals (to increase the probability of finding simultaneous events, and also to account for the possible difference in the internal clock of the computers). This value of 15~mn was chosen because it is shorter than the characteristic time of seeing fluctuations (discussed in section \ref{par:seeingtempcor}). That led to 901 triplets $S(h_1)$, $S(h_2)$ and $S(h_3)$ at altitudes $h_1=3$~m, $h_2=8$~m and $h_3=20$~m. Two different analyses were made from these data, giving similar results.

%
%________________________________________________________________

\subsubsection{First analysis}
The first calculation was made from the mean values $\bar{S_i}$ of $S(h_i)$. The 3 mean values for the 3 altitudes are  $\bar{S_1}=26.25\:10^{-13} m^{1/3}$, $\bar{S_2}=15.25\:10^{-13} m^{1/3}$ and $\bar{S_3}=7.65\:10^{-13} m^{1/3}$. 

As we are interested in the content of the SL only, we first subtracted the contribution of the ASL turbulence, estimated from the mean seeing of 0.36 arcsec deduced from the integral of curves a, i.a. $S(h_{sl}) = 1.38\, 10^{-13}$ m$^{1/3}$. This gives 3 values, denoted as $S_i$ which are: $S_1 = 24.87$, $S_2 = 13.87$ and $S_3 = 6.27$ in units of $10^{-13}$ m$^{1/3}$. These values average all measurements, they include situations when the telescopes are inside the SL and are above the SL. The latter does not provide information on the SL content. To estimate the mean value of $C_n^2$ inside the SL, we need the mean value of $S$ restricted to the situations when the telescope is located inside the SL. This is obtained by dividing the 3 values of $S_i$  by the corresponding probabilities $P(h_i)$ of being inside the SL.  Then we obtain 3 triplets $\tilde S_i = S_i /P(h_i)$. 

A simplified 3-level quantitative model of $C_n^2(h)$ inside the SL can then be estimated from 3 to 20m and can be compared to the average of the 32 samples provided by the radio soundings, after correction for the probability $P(h)$ for the same reason as above (the averaged profile at a given altitude contains a fraction of events outside the SL).  

As an integral is an additive process, the differences $\tilde S_1 - \tilde S_2$ and $\tilde S_2 - \tilde S_3$ provide the mean values of $C_n^2$ respectively between 3 and 8m and between 8 and 20m. The results are $\bar C_n^2=2.\, 10^{-13}\; m^{-2/3}$ between 3~m and 8~m and $\bar C_n^2=0.6\  10^{-13}\; m^{-2/3}$.  $\tilde S_3$ itself is the integral of $C_n^2$ from 20m to the top of the SL. The mean altitude of the SL when it is higher than 20m can be estimated from the radio soundings. This conditional mean is of the order of 40m, that indicates a mean value of $C_n^2$ in the altitude range $[20~m, h_{sl}]$ of $0.3\, 10^{-13}\; m^{-2/3}$. Fig. 10 shows the comparison of this rough, 3-layer profile with the one obtained from the balloon radio soundings. No information is available near the snow surface, below the lowest telescope. It would not make sense to extend the curves above 40m, where the statistical information is poor (the SL does not reach this altitude more than 10\% of the time). The two integrals up to 40m agree within 10\%, which is good, given the two very different technical approaches and the non simultaneity of the measurements. It is clear that the radio soundings do not provide reliable values below 20m. Il seems also clear from the DIMM data that the lowest layers are by far the most turbulent. Around 50\% of the turbulent energy of the SL is in the first 10 to 15 meters, and 95\% of the turbulent energy of the total atmosphere is inside the SL, which is more than previously thought.

\begin{figure}
\includegraphics[width=\figwidth]{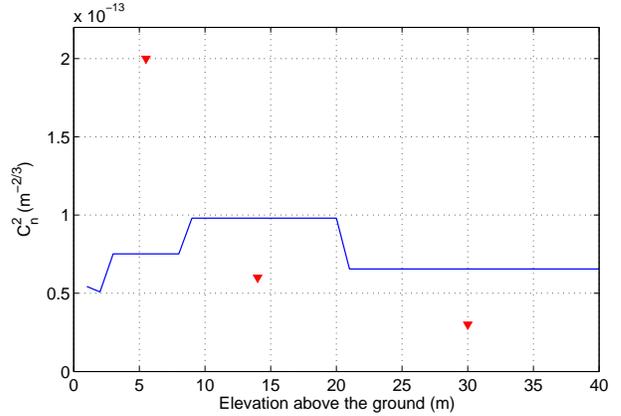}
\caption{The mean value of $C_n^2$ inside the turbulent SL, seen by the radio soundings (continuous curve) and modelled from the 3 DIMMs time series (triangles). The first few meters above the surface are poorly estimated. Given the different technologies and the different epochs of these measurements, the general agreement above 8~m is satisfactory. Values of $C_n^2$ at each altitude take in account the situations where the SL reaches at least this altitude (it is not an average including the nearly zero values when the SL is entirely below this altitude), as described in the text.}
\label{fig:blcontent}
\end{figure}

\begin{figure}
\includegraphics[width=\figwidth]{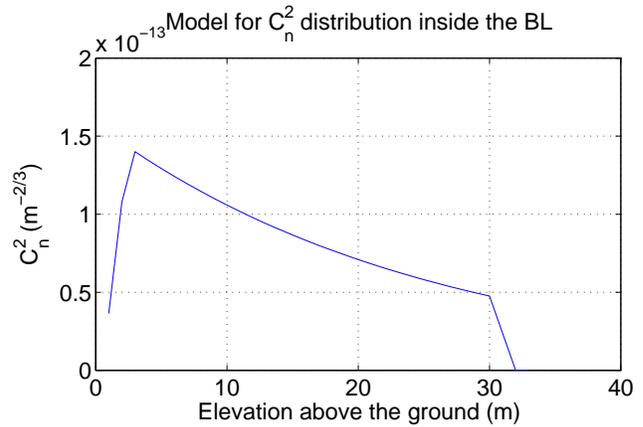}
\caption{A "typical" example of the $C_n^2$ SL. The upper limit is always extremely sharp.}
\label{fig:cn2blmodel}
\end{figure}

From these results, we can propose a description of the ``typical'' SL: starting at an altitude $h_b\simeq 2$~m, ending very sharply at a variable altitude $h_{sl}$ with an exponential probability, and with a content that can be deduced from both the poor seeing measurements and the winter balloon radio soundings. Fig \ref{fig:cn2blmodel} shows this typical SL. This figure is a ``typical'' example, valid at one given moment, and is not an average. Several radio soundings look very much like this model (flights number 544, 559, 565 or 568).

\subsubsection{Second analysis}
We made a second analysis on the integral triplets $S(h_1)$, $S(h_2)$ and $S(h_3)$. The idea is to compute instantaneous values of the SL thickness $h_{sl}$ for every occurrence of simultaneous measurements by the 3 DIMMs. 
 We modelled the function $C_n^2(h)$ by a piecewise function equal to a negative exponential law in the interval $h\in [h_b,\hsl]$ ($h_b$ is the lower limit
  of the surface layer) and 0 outside the interval (see Fig.~\ref{fig:cn2blmodel}). 
 The function $S(h)=\int_h^\infty C_n^2(h')\, dh'$ is then 
 \be
 S(h)=\left\{
 \begin{array}{ll}
 K & \mbox{if $h\le h_b$} \\
 \displaystyle K\; \exp (-a\: (h-h_b)) & \mbox{if $h_b\le h \le \hsl$} \\
 K_1 & \mbox{if $h\ge \hsl$} \\
\end{array}
 \right.
 \ee
With  $K_1=K\; \exp (-a\: (\hsl-h_b))$. $S(h)$ depends upon the four parameters $h_b$, $\hsl$, $a$ and $K$. The 3 measured values of $S(h)$ are not enough to estimate them without ambiguity. Moreover the situations where $h_3<\hsl$ do not allow us to estimate $\hsl$ by direct adjustment of the model to the data (in this case only a lower value of $\hsl$ could be given). We proceeded as follows: first, we made two hypotheses: (i) the two lowest DIMMs were always inside the SL (i.e. $h_b\le h_{1,2} \le \hsl$) and (ii) the SL contains 95\% of the total turbulent energy, as stated previously. The first hypothesis was checked from the seeing measured by the two lowest DIMMS (a seeing of the order of 0.3\arcsec means that the DIMM is outside the SL, or that the SL did not exist at this particular moment). The hypothesis appeared to be valid in 94\% of the cases, and the following computation was made on these cases.

\begin{figure}
\includegraphics[width=\figwidth]{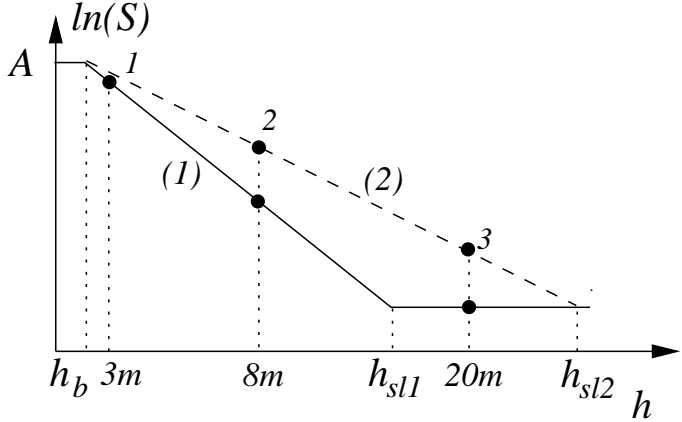}
\caption{Piecewise model for $S(h)$. Points 1, 2 and 3 correspond to the DIMMs at elevations $h_1$, $h_2$ and $h_3$. Two situations are shown. (1) the SL upper limit $h_{sl1}$ is lower than $h_3$ and the 20-m DIMM is outsite the SL. (2) the SL upper limit is higher than $h_3$ and the point 3 is inside the SL. Points 1 and 2 are supposed to be always inside the SL.}
\label{fig:geom_sl}
\end{figure}

The roof-based DIMM could be either outside or inside the SL (as illustrated by Fig.~\ref{fig:geom_sl}). The latter corresponds to a function ln$(S(h))$ showing 3 aligned points with a slope $-a$ which was computed by least-square fitting on the 3 points. Otherwise, the slope was computed using only the points 1 and 2. The value of $\hsl$ was computed so that $S(\hsl)=0.05\: S(h_b)$ (i.e. the turbulent energy above the SL represents 5\% of the turbulent energy above $h_b$).

\begin{figure}
\parbox[t]{6cm}{\hskip -3mm\includegraphics[width=65mm]{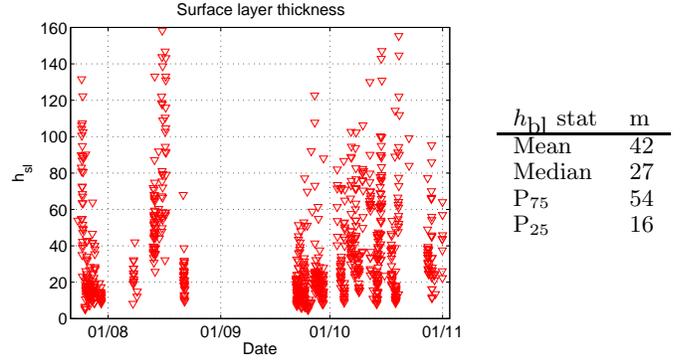}} \ \parbox{2cm}{\vskip -5cm
\begin{tabular}{ll}
$h_{\mbox{bl}} $  stat & m\\ \hline
Mean   & 42\\
Median & 27\\
P$_{75}$ & 54\\
P$_{25}$ & 16\\
\end{tabular}}
\caption{SL thickness $h_{sl}$ as a function of time for each seeing triplet. Corresponding statistics are displayed in the table on the right.}
\label{fig:hslvsdate}
\end{figure}

This was applied to all the seeing triplets and led to a mean value $h_{sl}=42$~m and a median of $27$~m, which is a little more pessimistic that our previous analysis. The time series of $h_{sl}$ and its statistics are shown in Fig.~\ref{fig:hslvsdate}.

 %
 %------------------------------

\subsection{The ASL seeing}
\label{par:faseeing}
\begin{figure}
\parbox[t]{5cm}{\includegraphics[width=55mm]{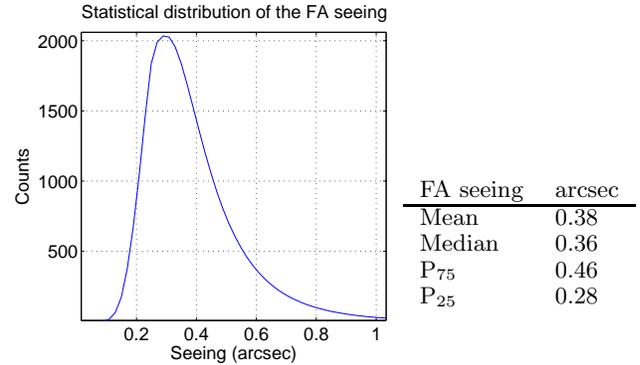}} \ \parbox{2cm}{\vskip -3cm
\begin{tabular}{ll}
FA seeing& arcsec\\ \hline
Mean & 0.38\\
Median & 0.36\\
P$_{75}$ & 0.46\\
P$_{25}$ & 0.28\\
\end{tabular}}
\caption{Distribution of the ASL seeing at Concordia, measured during 10 seasons between 2005 and 2008, and excluding summers. The table on the right the corresponding statistics.}
\label{fig:histoseeingfa}
\end{figure}

The values of the left peaks of all seeing histograms (excluding summers) can be summed to provide a PDF of the ASL seeing that proves to be, within our statistics, independent of the season and of the altitude in the first 20 meters,  provided that the SL is completely below this altitude. Figure \ref{fig:histoseeingfa} shows the sum of 10 seasons (3 winters, 3 springs and 4 autums). It can be regarded as a very robust estimation of the ASL PDF of the seeing, especially since it does not appear to depend on the season. We know that the high altitude winds increases in winter and reduce, for instance, the isoplanatic angle, but they do not significantly influence the seeing. Even the very peculiar case of summer (see section \ref{par:seeingete}) shows a mean seeing of the order of 0.3 arcsec at 5 p.m. local time, when the SL essentially disappears, so that the ASL seeing becomes accessible every day for a moment at surface level. This PDF, with a log-normal distribution, is characterized by two key numbers: seeing($P_{max}$) = 0.29 arcsec, and median = 0.36 arcsec. Further statistics are displayed in Figure \ref{fig:histoseeingfa}.

%
% --------------------------------------
\subsection{The peculiar case of the summer seeing}
\label{par:seeingete}
\begin{figure}
\parbox[t]{5cm}{\includegraphics[width=55mm]{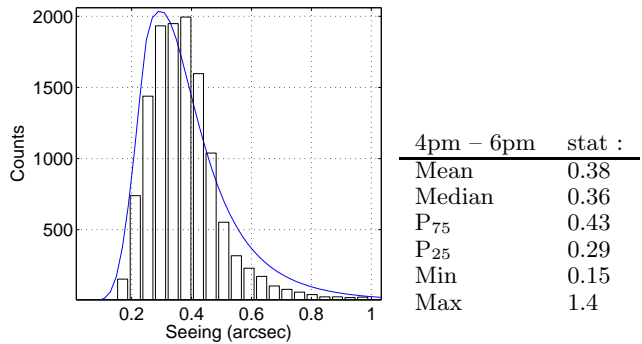}} \ \parbox{2cm}{\vskip -3cm
\begin{tabular}{ll}
4pm -- 6pm & stat :\\ \hline
Mean & 0.38\\
Median & 0.36\\
P$_{75}$ & 0.43\\
P$_{25}$ & 0.29\\
Min & 0.15\\
Max & 1.4\\
\end{tabular}}
\caption{Histogram and statistics of the seeing values corresponding to the period 4pm-6pm in summer (December and January). The superimposed solid curve is the ASL seeing distribution.}
\label{fig:histosfa}
\end{figure}

In summer, the SL becomes much weaker as both the temperature and the wind speed gradients are significantly weaker than in cold seasons. As already mentioned several times (i.e. Aristidi et al. 2005), they even disappear every day at 5 p.m, while these gradients are similar to the winter conditions at local midnights. This seeing depending strongly on local time does not permit the histograms to discriminate different situations that are not sharply separated. The summer histograms tend to show a unique log normal distribution (Fig.~\ref{fig:histosfa}), with a mean seeing much better than in other seasons but with the ASL seeing being lost in the mixing of the different situations. Only the daily average at 5 p.m permits us to confirm that this ASL seeing is, again, the same as in all other seasons. 

We thus selected the seeing data (at elevation 8 m) measured in summer (we selected only the months of December and January, the closest to the summer solstice) in the local time range 4pm--6pm, i.e. a 2 hour interval in which the seeing is usually the best. Results are shown in Figure \ref{fig:histosfa}. The statistics of such events is indeed very close to the FA distribution presented above, the mean and the median value being identical. 

The same analysis made on the 3 m DIMM data shows a minimum median value of the seeing of 0.57 arcsec at 5~pm. This is much greater than the FA median seeing of 0.36 arcsec, and would indicate that the SL does not totally disappear and that there is still a little turbulence between 3 and 8 m that may be caused by surface effects.

%
%________________________________________________________________

\subsection{Temporal fluctuations of the seeing}
\label{par:seeingtempcor}
This is an important point to consider high angular resolution imaging. Temporal fluctuations of the seeing were studied by Racine~(1996) and by Ziad et al.~(1999) and gave a characteristic time of 17~minutes for the two temperate sites of Mauna Kea and La Silla.
At Dome C, we noticed that the periods of ``good'' seeing (0.3 arcsec or lower) are shorter than the periods of ``bad'' seeing ($>$ 1 arcsec) and we propose here a different approach to estimate the characteristic time of seeing fluctuations as a function of the seeing. It is quite a difficult question as we lack very long sequences of uninterrupted data sets. For various technical reasons, our  data sequences have interruptions that make this specific statistical study somewhat inaccurate.

We define here an interval of stability as a continuous period of time in which the seeing is less than a given threshold $s_0$. We denote as $t_s$ its length. In this interval, we allow the seeing to be greater than or equal to $s_0$ during 10\% of $t_s$. For example if the seeing is less than 0.5 arcsec for one hour with a small interruption of 5 minutes, this interruption is neglected and $t_s$ will be set to one hour. For a given value of $s_0$, the histogram of $t_s$ shows a negative exponential behaviour whose mean is taken as the characteristic time of the seeing stability (for an exponential distribution, the mean is the 63\% level). The minimum possible value of $t_s$ is 2 minutes, which is the temporal sampling of the DIMMs. Events correspondigs to $t_s$=2~mn occur quite often when the seeing fluctuations are large or when $s_0$ is close to the actual seeing value. They tend to bias the calculation of the mean of $t_s$ and are thus neglected.

\begin{figure}
\includegraphics[width=\figwidth]{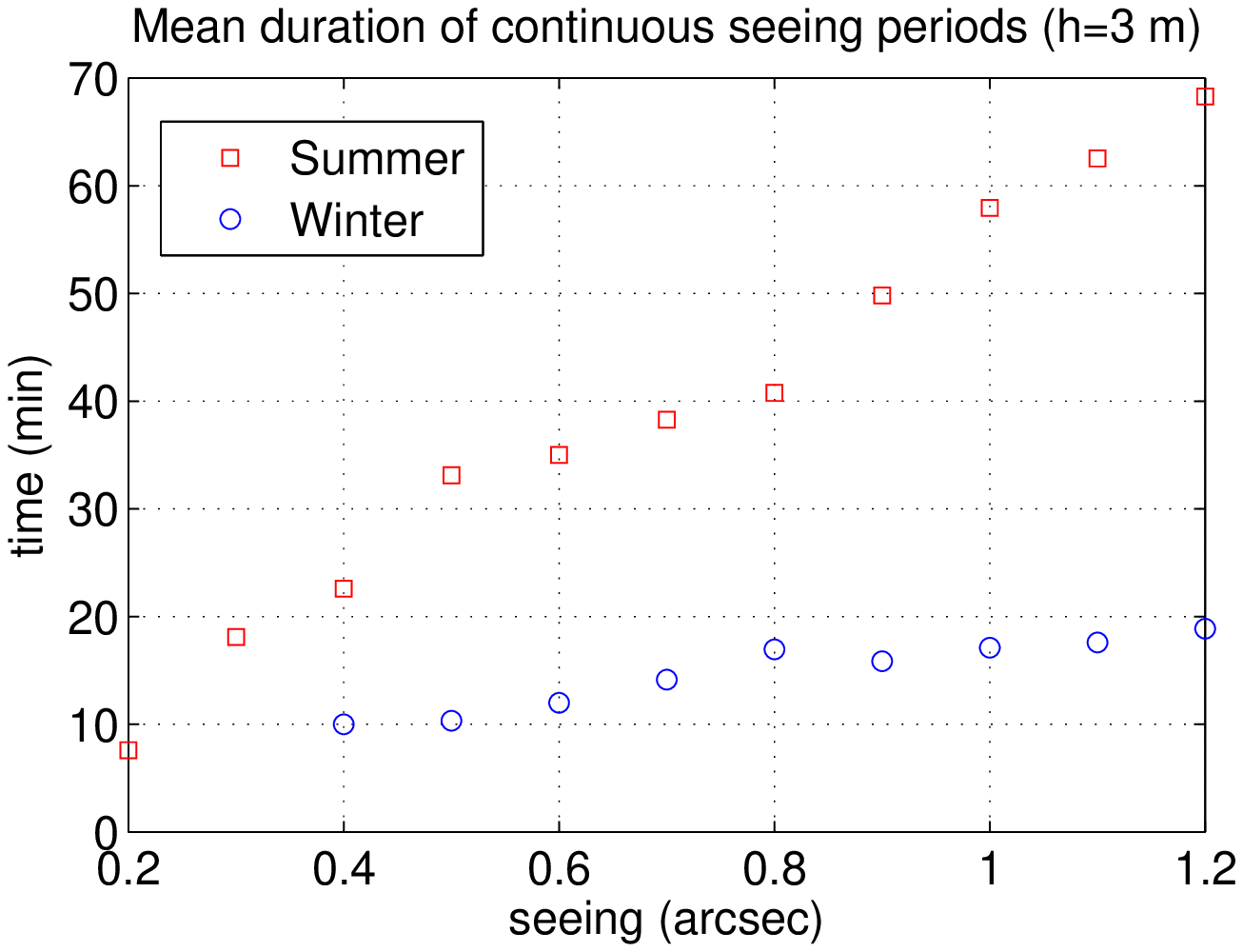} 
\caption{Histogram of the length $t_s$ of continuous intervals with seeing $< s_0$, as a function of $s_0$, for seeing data taken in summer and in winter winters with the DIMM at an elevation of 3~m.}
\label{fig:ts_3m}
\end{figure}

\begin{figure}
\includegraphics[width=\figwidth]{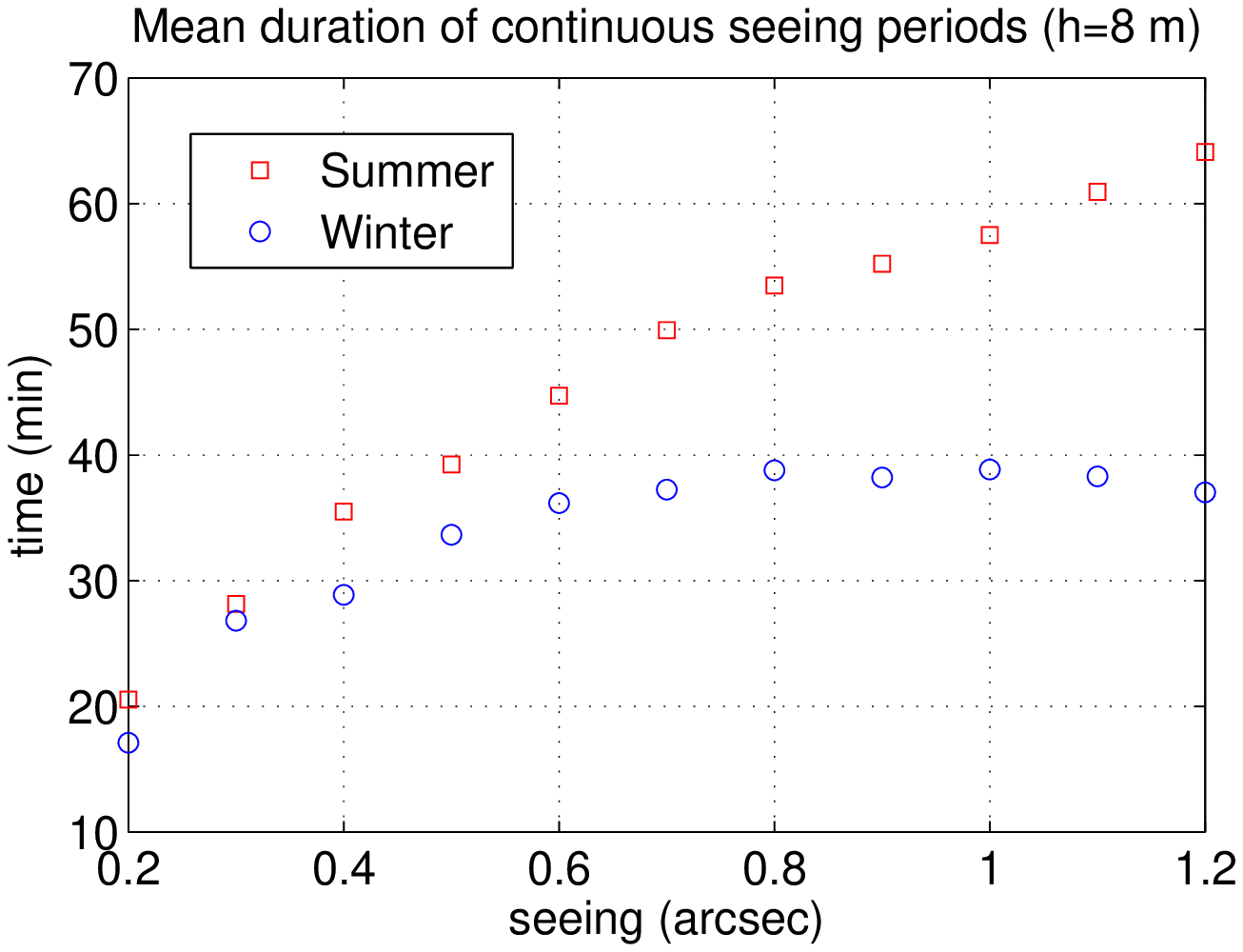} 
\caption{Histogram of the length $t_s$ of continuous intervals with seeing $< s_0$, as a function of $s_0$, for seeing data taken in summer and in winter with the DIMM at an elevation of 8~m.}
\label{fig:ts_8m}
\end{figure}

We performed this analysis for the DIMMs located at elevations 8~m and 3~m for the summer and the winter periods. The result is displayed in Figs~\ref{fig:ts_3m} and~\ref{fig:ts_8m}. One can see that the characteristic time increases with the seeing and saturates. It is of the order of half an hour in winter for seeing values around 0.5 arcsec at elevation 8 m, while it is only 10 minutes at an elevation of 3 m.

The curves are likely to be polluted by the interruptions of the observations. The ``up-time'' intervals have a mean of 90~min in summer and 82~min in winter for the DIMM at 8 m. It is 199 min in summer and 138 min in winter for the DIMM at 3 m. The saturation of the curves is due both to the fact that the seeing has a maximum value (and the saturation is observed when the threshold becomes of the order of this maximum value), and due to the interruptions. 

\begin{figure}
\includegraphics[width=\figwidth]{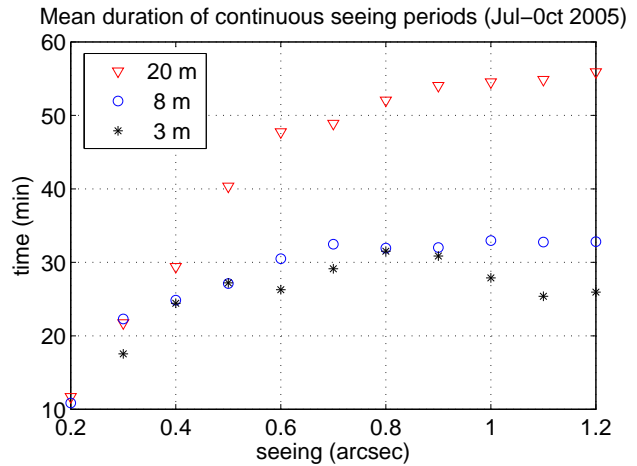} 
\caption{Histogram of the length $t_s$ of continuous intervals with seeing $< s_0$, as a function of $s_0$, for seeing data taken at the three elevations between July and October 2005.}
\label{fig:ts_3niv}
\end{figure}

We did the same analysis for the seeing data taken by the DIMM on the roof of the base. Since the covered period is much shorter (100 days between July and October 2005), we present in Fig.~\ref{fig:ts_3niv} the curves of $t_s$ versus $s_0$ for this period of time for the 3 heights (3~m, 8~m and 20~m). This graph shows that the characteristic time is longer at the height of 20~m (the gain is more than 50\% for a seeing $s_0=0.5$ arcsec).

%

%
% --------------------------------------
\subsection{Isoplanatic angle}
Direct measurements of the isoplanatic angle $\theta_0$ can be obtained by observing the scintillation of a single star through an aperture of diameter 10~cm with a central obstruction of 4~cm (Loos \& Hogge 1979). The scintillation index $\sigma^2$ (intensity variance normalized by the square of the mean intensity) of the star at a zenithal angle $z$ can be related to the isoplanatic angle by 
\be
\theta_0^{-5/3}\: =\: A\: \cos(z)^{-8/3}\: \sigma^2
\ee
with $A=0.1963$. We used for these measurements a telescope of GSM during GSM down time. Details of this so-called ``DIMM-$\theta_0$'' monitor can be found in Aristidi et al. (2005). Observations were made in the visible domain ($\lambda\in [320, 630]$ nm) with exposure times of 5~ms and 10~ms. A bias correction for the exposure time was applied by linear extrapolation on the scintillation indexes as described in Sect. 4.2 of  
Aristidi et al. (2005).

A total of 46653 values of $\theta_0$ were obtained. They are sparsely distributed over the period 2004--2006; data are available for the months of January 2004, May to July 2005 and January to May 2006. Statistics and the histogram of the distribution are shown in Fig.~\ref{fig:histoisop}. We found a median value of 3.9\arcsec, which is slightly lower than the 5.7\arcsec\ found by Lawrence et al. (2004) with MASS but still very competitive compared to more classical sites (Table 1 of Lawrence et al. 2004). Our value is very close to the South Pole isoplanatic angle of 3.2\arcsec\ as estimated by Marks et al. (1999) from in situ radio soundings. This is not surprising; the isoplanatic angle expressed as a weighted integral over the vertical distribution of turbulence is
\be
\theta_0^{-5/3} \propto \int_0^\infty h^{5/3}\, C_n^2(h)\, dh
\ee
the weight $h^{5/3}$ gives more sensitivity to the high altitude turbulence. Both sites of Dome C and South Pole actually exhibit strong winds in winter above 10 km (Marks et al. 1999; Trinquet et al. 2008a).

\begin{figure}
\includegraphics[width=\figwidth]{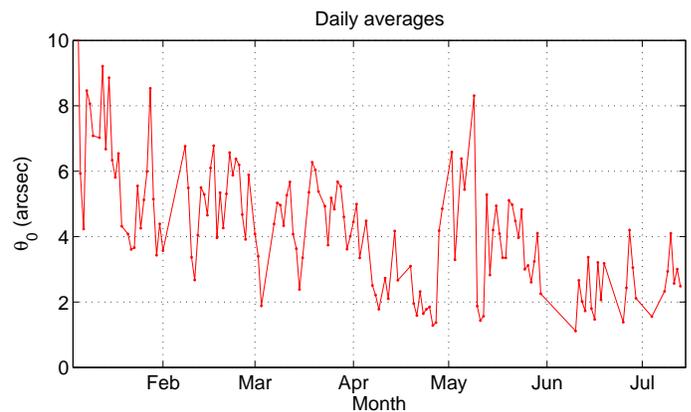} 
\caption{Isoplanatic angle $\theta_0$ as a function of the month.}
\label{fig:isopvsmonth}
\end{figure}

\begin{figure}
\parbox[t]{6cm}{\hskip -3mm\includegraphics[width=65mm]{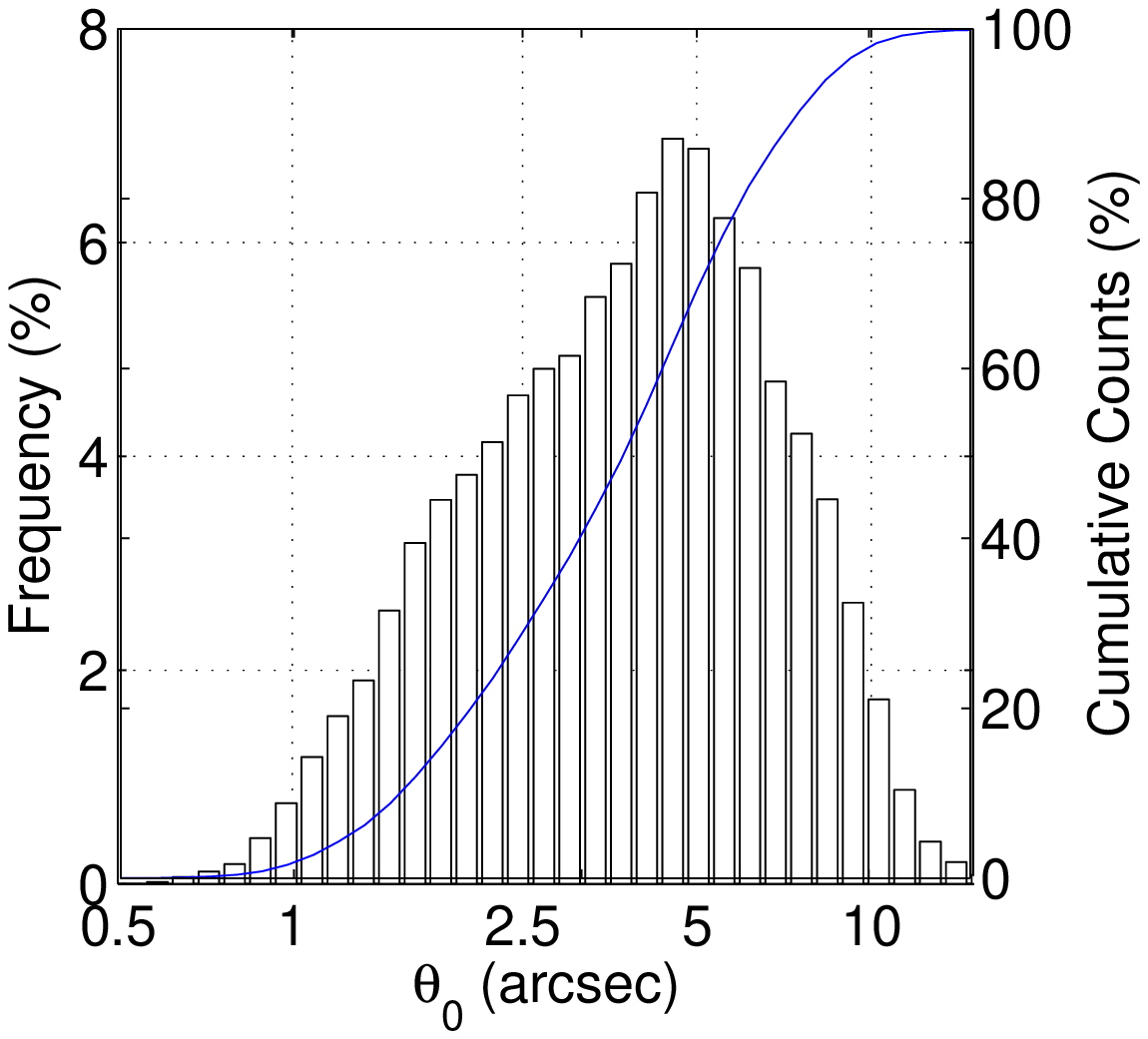}} \ \parbox{2cm}{\vskip -5cm
\begin{tabular}{ll}
$\theta_0 $  stat & arcsec\\ \hline
Mean   & 4.3\\
Median & 3.9\\
P$_{75}$ & 5.7\\
P$_{25}$ & 2.4\\
Min    & 0.4\\
Max    & 23.4\\
\end{tabular}}
\caption{Isoplanatic angle histograms and statistics. Values are in arcsec.}
\label{fig:histoisop}
\end{figure}

Figure~\ref{fig:isopvsmonth} shows the dependence of $\theta_0$ on the period of the year. Data are daily averaged and plotted as a function of the month regardless of the year. We see that $\theta_0$ is better in summer (about 7\arcsec), and decreases to 3\arcsec\ in winter. This is consistent with the value of 2.7\arcsec\  published by Trinquet et al. (2008a) estimated from radio soundings.

\begin{figure}
\includegraphics[width=\figwidth]{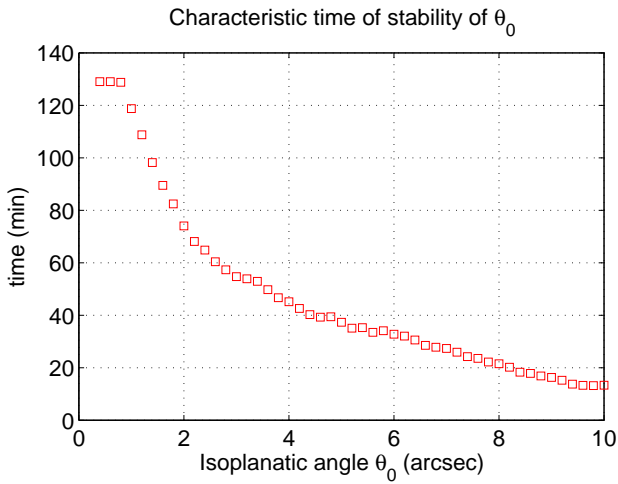} 
\caption{Mean duration of intervals in which the isoplanatic angle is greater than a given value $\theta_0$.}
\label{fig:isoptempcor}
\end{figure}

Temporal fluctuations of $\theta_0$ were investigated as well. The method described in section~\ref{par:seeingtempcor} was applied to compute the mean time where the isoplanatic angle is greater than a given value $\Theta_0$. The curve displayed in Figure~\ref{fig:isoptempcor} unsurprisingly shows a negative slope with a saturation of the first points (i.e. values below 1 arcsec) due to the gaps in the data. We can see that the periods where $\theta_0$ is greater than 5 arcsec last 40 minutes on average.

%
%--------------------------------------
%
\section{Conclusion}
We have presented statistics of the seeing and isoplanatic angle at Dome C using all the DIMM data collected from the beginning of our observations about 4 years ago. Previous results were published in Aristidi et al. (2005) for the summer data and  Agabi et al. (2006) for the first half of the first winter. We confirm here the general trend from these first data, i.e. that the seeing is exceptional in summer and can reach values of 0.3 arcsec at a few meters above the ground, and that it degrades to very poor values of around 2 arcsec in winter. The cause is the occurrence of a very strong turbulent layer near the surface that appears in winter. This situation was reproduced 3 times, leading to a sinusoide like curve of the seeing as a function of time. The isoplanatic angle, first found to be excellent in summer (median value of 7 arcsec) also degrades in winter, but not for the same reason. It has been shown that the wind speed increases at high altitude in the free atmosphere in winter (Trinquet et al. 2008a; Geissler \& Masciadri 2006) which creates high altitude turbulence and divides the isoplanatic angle by a factor of 2 in winter, although the seeing itself does not display significant variations. 

The temporal stability of the seeing and the isoplanatic angle at Dome C have been adressed for the first time, and provide answers to key questions. We found (at 8 m) that the characteristic time for the seeing being lower than a given threshold is around 30~minutes if this threshold is 0.5 arcsec, i.e. for the periods of ``good'' seeing. The same order of magnitude holds for the isoplanatic angle, since the characteristic time of stability for values higher than 5 arcsec is also around 30~mn.

Using DIMMs at three different elevations, and through the very special geometry of the SL, it is possible to discriminate clearly  the ASL from the SL seeing and to study the statistical properties of both. This is a very remarkable result for an instrument as simple as the DIMM. The SL appears to be sharply defined between two heights, starting 2~m above the ground and ending at a median height of 23 to 27~m. Its profile $C_n^2(h)$ is exponential-like, and a telescope located 23~m above the ground will be above the SL half of the time. This is the first time that such quantitative results have been obtained, however based on only the 100 days of data (July to October 2005) when the three DIMM were operated together. This emphasizes the need for more statistics, requiring a new DIMM on the roof of the base (or on the top of the 45 m mast, but that is clearly a more difficult technical task) for a longer period of time (at least one full year).

The seeing above the surface layer is, as expected, excellent, while slightly less to the values published by the Australian group operating a MASS+SODAR combination in 2004 (Lawrence et al. 2004). These ASL conditions are accessible in summer at around 5~pm every day for a temporal window of roughly 2 hours, when the statistics of the seeing at an elevation of 8~m is almost the same as the winter ASL seeing. This is of course very interesting for solar observations as stated before (Arnaud et al. 2007) but some turbulence does still exist in the first meters above the ground even at 5~pm in the summer and that a solar telescope must be located at least at 8~m to benefit from the excellent seeing of the FA.

Finally, the turbulent profile inside the SL has also been deduced from these DIMM data sets by exploiting only the fraction of time spent by these DIMMs inside the turbulence. It is remarkably consistent with the few samples directly measured in-situ by the radio soundings of the winter balloons, and an important point to be noted is that 95 percent of the total turbulence energy above 3 m is inside the SL.

During the polar summers 2006-2007 and 2007-2008 a set of 6 sonic anemometers was installed on the 45~m mast to perform in situ monitoring of the temperature, wind speed and $C_n^2$ inside the BL (Travouillon et al., 2008). A part of the data corresponding to the winter 2007 are under analysing, but in 2007 the mast was only 33~m high and there were 3 sonic anemometers. The complete data for the 6 sonics have not yet been analysed. They will permit us to obtain better statistics and to address issues such as the relationship between the turbulent layer height and content, and meteorological parameters such as wind speed and temperature gradients.

%
%________________________________________________________________

\begin{acknowledgements} 
We wish to thank the French and Italian Polar Institutes, IPEV and PNRA and the CNRS for the financial and logistical support of this programme. Thanks also to the different logistics teams on the site who helped us to set up the experiments. We are also grateful to our industrial partners ``Optique et Vision'' and ``Astro-Physics'' for technical improvements of the telescopes and their help in finding solutions to critical technical failures on the site. Thanks to our colleagues Max Azouit and Fran\c{c}ois Martin for constant help and ideas since the beginning of the project. Thanks to F. Vakili, the director of Fizeau laboratory, for his support. We would also like to cite the people who participated in the summer campaigns: Merieme Chadid, Jean-Baptiste Daban, Tatyana Sadibekova, Caroline Santamaria, Fran\c{c}ois-Xavier Schmider, Tony Travouillon and Franck Valbousquet.  Finally, we thank Nicolas Epchtein for his critical reading of the manuscript.
\end{acknowledgements}

\end{document}